\documentclass[reprint,aps,prl,twocolumn ,groupedaddress,nobibnotes]{revtex4-1}

\usepackage{amsmath,amsfonts,amssymb,graphicx,bbm,times}
\usepackage{dsfont}
\usepackage{listings}
\usepackage[usenames,dvipsnames,table,xcdraw]{xcolor}   
\usepackage{hyperref}
\usepackage{lipsum}
\usepackage{cleveref} 														
\usepackage{dcolumn}   														
\usepackage{cancel}
\usepackage{soul}

\DeclareMathOperator{\Tr}{Tr}

\hypersetup{colorlinks=true,
						linkcolor=blue,      
						citecolor=blue,            
						filecolor=blue,             
						urlcolor=cyan              
}
\begin{document}

\preprint{\bf PREPRINT}

\newcommand{\id}{{\mathbbm{1}}}
\renewcommand{\vec}[1]{\boldsymbol{#1}}

\def\>{\rangle}
\def\<{\langle}
\def\({\left(}
\def\){\right)}

\newcommand{\ket}[1]{\left|#1\right>}
\newcommand{\bra}[1]{\left<#1\right|}
\newcommand{\braket}[2]{\<#1|#2\>}
\newcommand{\ketbra}[2]{\left|#1\right>\!\left<#2\right|}
\newcommand{\proj}[1]{|#1\>\!\<#1|}
\newcommand{\avg}[1]{\< #1 \>}
\renewcommand{\tensor}{\otimes}
\newcommand{\einfuegen}[1]{\textcolor{PineGreen}{#1}}
\newcommand{\streichen}[1]{\textcolor{red}{\sout{#1}}}
\newcommand{\todo}[1]{\textcolor{blue}{(ToDo: #1)}}
\newcommand{\transpose}[1]{{#1}^t}
\newcommand{\om}[1]{\textcolor{red}{#1}}

\title{Dissipation-assisted matrix product factorization}
\author{Alejandro D. Somoza, Oliver Marty, James Lim, Susana F. Huelga, Martin B. Plenio}
\affiliation{
Institut f\"ur Theoretische Physik and IQST, Albert-Einstein-Allee 11, Universit\"at Ulm, D-89069 Ulm, Germany
}

\begin{abstract}
Charge and energy transfer in biological and synthetic organic materials are {strongly influenced} by the coupling of electronic states to high-frequency underdamped vibrations under dephasing noise. Non-perturbative simulations of these systems require a substantial computational effort and current methods {can only be} applied to large systems with severely coarse-grained environmental structures. In this work, we introduce a dissipation-assisted matrix product factorization (DAMPF) method based on a memory-efficient matrix product operator (MPO) representation of the {vibronic state at finite temperature}. In this approach, the correlations between {environmental} vibrational modes can be controlled by the {MPO bond dimension}, {allowing for systematic interpolation} between approximate and numerically exact dynamics. Crucially, by subjecting the vibrational modes to damping, we show that one can significantly reduce the bond dimension required to achieve a desired accuracy, and also consider a continuous, highly structured spectral density in a non-perturbative manner. We demonstrate that our method can simulate large vibronic systems consisting of 10-50 sites coupled {with 100-1000 underdamped modes in total} and for a wide range of parameter regimes. {An analytical error bound is provided} which allows one to monitor the accuracy of the numerical results.  This formalism will facilitate the investigation of {spatially extended} systems with applications to quantum biology, organic photovoltaics and quantum thermodynamics. 
\end{abstract}
\date{\today}
\maketitle



\textit{Introduction}---{Vibronic networks describe spatially extended systems of electronically interacting sites coupled to underdamped intramolecular vibrational modes, thus providing a general framework for the description of energy and charge transfer processes. Relevant examples include natural and artificial photosynthetic pigment-protein complexes \cite{Valkunas_book,Blankenship2002book,Scholes2017} and organic photovoltaics \cite{Antonietta2017}. The role of vibronic coupling has also been emphasized in the context of singlet fission \cite{Stern2017}, charge separation \cite{Romero2014,Fuller2014,Falke2014} and polaron formation \cite{DeSio2016} at different donor-acceptor interfaces. Modelling realistic dissipative effects requires the electronic states and intramolecular vibrations in the network to interact with bosonic reservoirs at finite temperatures, making simulations of spatially extended vibronic systems particularly challenging. However, the ability to perform this type of simulations is crucial to make further progress in the understanding of transfer phenomena at the microscopic level and, in particular, discerning the possible relevance of coherent effects in the actual dynamics of these complexes \cite{Huelga2013,Chin2013,Lim2015}.}

\begin{figure}
	\includegraphics[height=0.5\textwidth]{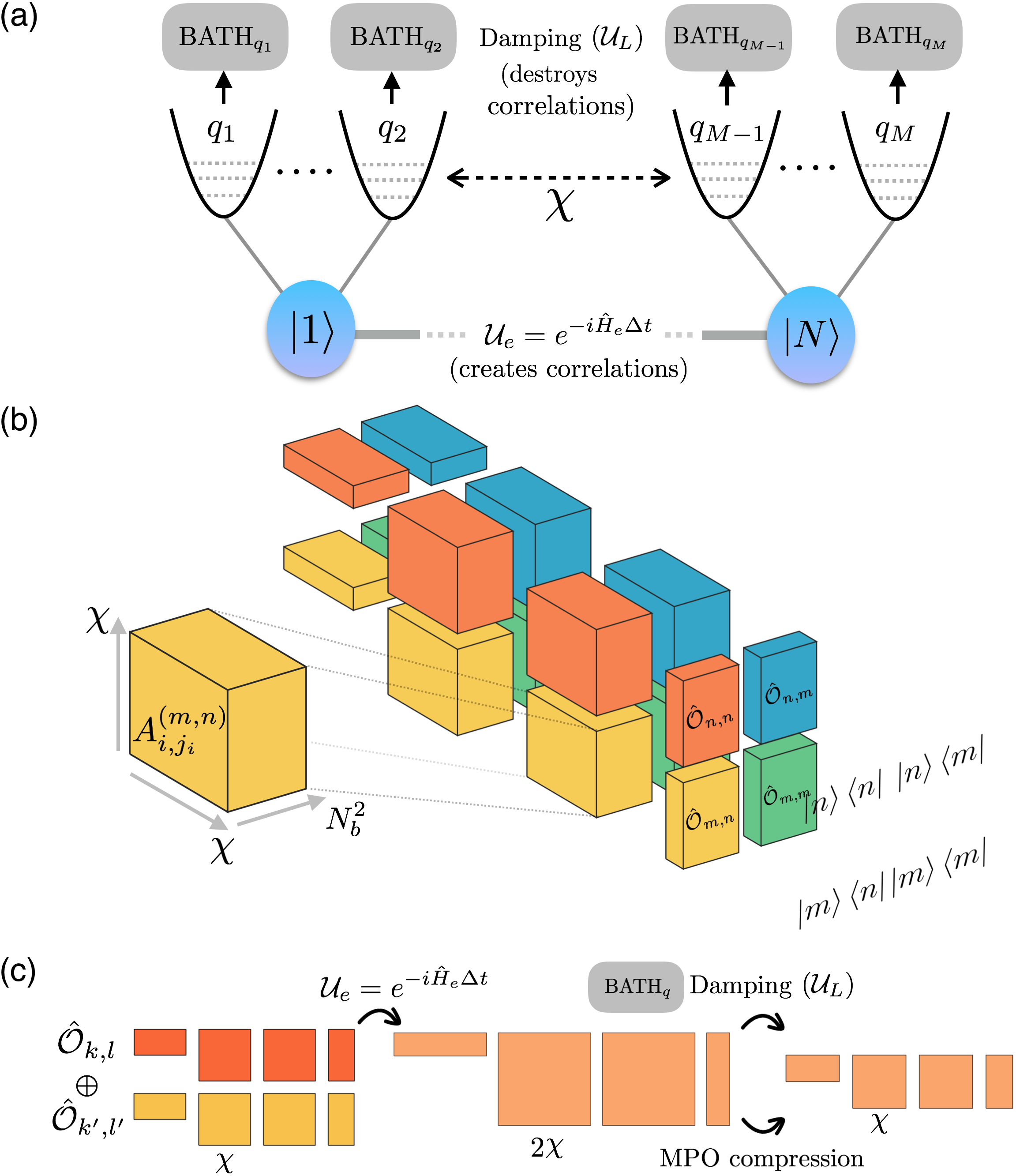} 	
	\caption{(a) Schematic of a vibronic network consisting of $N$ electronically interacting sites where each site is locally coupled to $Q$ oscillators under Lindblad damping. $M=NQ$ denotes the total number of the modes. A bond dimension $\chi$ determines how accurately the correlations between oscillators are described using MPOs. (b) Illustration of the MPO representation of the quantum state of a vibronic system. For every operator $\ket{m}\bra{n}$~ of the electronic subsystem there is a MPO describing the conditional state $\hat{\mathcal{O}}_{m,n}$ of the nuclear degrees of freedom (Eqs.(\ref{eq:rho}-\ref{eq:OnmMPO})). $N_b$ is the number of oscillator levels employed to describe each vibrational mode. { (c) The electronic interaction $\mathcal{U}_e$ creates correlations between oscillators, mediated by the sum of $\hat{\mathcal{O}}_{k,l}$ and $\hat{\mathcal{O}}_{k^\prime,l^\prime}$ (see Eq.(\ref{eq:electronic_update})). This sum doubles the bond dimension and a compression scheme is required. Crucially, the local environment at each site ($\mathcal{U}_L$, Eq.(\ref{eq:damping})) suppresses these correlations and helps one to keep the bond dimension under control.}} 	\label{fig:model_MPOrepresentation}
\end{figure}

The comparable coupling strengths amongst electronic states and nuclear coordinates in organic materials can lead to highly non-Markovian dynamics \cite{Rivas2014,Fruchtman2015,Iles-Smith2016,Strathearn2018} where perturbative {master equations such as Redfield theory \cite{Redfield1957}} are inaccurate. {Numerically exact simulations} of the quantum dynamics of vibronic systems are often restricted to only a few sites due to the rapid growth of the Hilbert space with system size, and {as a result}, vibrational environmental structures need to be severely coarse-grained in order to study large systems. In this work, we approach this challenge by employing matrix product operators (MPOs) that allow us to represent the state of the vibrational environment in a memory-efficient manner. { This approach filters some quantum correlations between the modes by controlling {a parameter}, the bond dimension of the MPOs}. In spite of the benefit of this representation, these correlations generally grow in time and require increasingly high bond dimensions. We {address} this issue by splitting the system into a non-Markovian vibronic core which, in turn, is coupled to a secondary Markovian environment (see Fig.\ref{fig:model_MPOrepresentation}(a)). The benefit of this approach is two-fold: On the one hand, correlations between oscillators are dynamically suppressed by the secondary baths at finite temperatures. On the other hand, it enables the simulation of any continuous, highly structured spectral density in a {numerically} exact manner, based on the mathematical equivalence between reduced electronic dynamics under a continuous bosonic environment and that of a discrete environment with Lindblad damping \cite{Tamascelli2018a}. {Our} method can thus be employed to mimic tailored complex environments employing a few damped oscillators attached at each site. Such a pseudomode theory has been widely employed in the study of electron transfer in biomolecules \cite{Garg1985a},  the simulation of non-Markovian dynamics in open quantum systems \cite{Imamoglu1994,Martinazzo2011JCP} {and} the non-perturbative decay of atomic systems in cavities \cite{Garraway1997a,Dalton2001}, although its application to extended systems in combination with matrix product techniques has, {to the best of our knowledge}, not been proposed until now. 

{The main limitations of numerically exact methods addressing vibronic systems are the accesible size of the system (number of sites), a faithful reproduction of the environmental {spectral density} and the dissipative evolution that leads to mixed quantum states. Some selected numerically exact methods {aimed at addressing these challenges} are the hierarchical equations of motion (HEOM) \cite{Tanimura1989,Tanimura2006,Ishizaki2009} and recent stochastic extensions \cite{Ke2017}, the multi-layer multi-configuration time dependent Hartree (ML-MCTDH) method \cite{Meyer1990}, the quasi-adiabatic propagator path-integral (QUAPI) method \cite{Makri1995} and time dependent density matrix renormalization group (TD-DMRG) methods in combination with the theory of orthogonal polynomials (TEDOPA) \cite{Prior2010,Chin2010,Chin2013,Woods2015a,Tamascelli2018b}. HEOM is limited both by memory and computation time when the vibronic interactions are strong and the modes are underdamped, especially in the presence of multiple modes. The ML-MCTDH and its recent Gaussian-based extensions \cite{Romer2013,Hughes2014,Eisenbrandt2018a} enables one to study systems consisting of dozens of electronic states and underdamped modes, but it is based on pure state evolution and requires statistical sampling to take into account finite temperature effects, which can be challenging at high temperatures \cite{Matzkies1999,Manthe2001}. QUAPI is limited when system-environment correlations are long-lived, which is the case for underdamped vibrational environments {and for low temperatures}. TEDOPA can accurately simulate dimers coupled to highly structured environments but generalization to multiple sites appears computationally challenging. Finally, a recent TD-DMRG algorithm \cite{Ren2018} approximates the full vibronic propagator as a MPO while the state is represented by a matrix product state (MPS). Although the method can be applied to spatially extended systems, the bond dimensions that are required for convergence can be very large due to the absence of damping of the oscillators}. In this work, we introduce a method { that addresses challenges of existing methods and} is able to simulate { efficiently} composite vibronic systems with highly structured environments.

\textit{Model}---The vibronic system is modeled by a network of $N$ sites with inter-site couplings $J_{m,n}$. Each site is locally coupled to $Q$ harmonic oscillators {each subject to} Lindblad damping, where $M=NQ$ is the total number of oscillators (Fig.\ref{fig:model_MPOrepresentation}(a)). We will consider the single-excitation manifold for the computation of energy transfer and linear optical responses, with $\ket{n}$ representing a local electronic excitation at site $n${, although an extension to multiple excitations is straightforward in our approach}. The electronic Hamiltonian is described by
\begin{align}
\label{eq:electronic_hamiltonian}
\hat H_e &= \sum_{n=1}^N \Omega_n \ketbra{n}{n} + \sum_{m \neq n} J_{m,n} \ketbra{m}{n}, 
\end{align}
where $\Omega_n$ denotes the {excitation} energy of site $n$. The vibrational Hamiltonian is given by $\hat H_v = \sum_{n=1}^N  \sum_{q=1}^Q \omega_{q} \hat a_{n,q}^\dag \hat a_{n,q}$ and the interaction between electronic and vibrational degrees of freedom is modeled by $\hat H_{ev} = \sum_{n=1}^N\sum_{q=1}^Q \omega_{q} \sqrt{s_q} \ketbra{n}{n}  \left(\hat a_{n,q} + \hat a_ {n,q}^\dag \right)$ where $\omega_{q}$ is the vibrational frequency of mode $(n,q)$ locally coupled to site $n$. $\hat a_{n,q}$ and $\hat a^\dag_{n,q}$ are the annihilation and creation operators, respectively, of the mode with the Huang-Rhys factor $s_{q}$ quantifying vibronic coupling strength. In addition to the Hamiltonian dynamics, {the modes are coupled to finite temperature thermal reservoirs described by Lindblad damping,}
\begin{align}
\label{eq:damping}
{\mathcal{U}_L\hat{\rho}}=&
\sum_{n,q} 2\gamma_q \bigg[ \bar n(\omega_{q}) \left( \hat{a}_{n,q}^\dag~\hat{\rho}~\hat{a}_{n,q} -\frac{1}{2}\lbrace \hat{a}_{n,q} \hat{a}_{n,q}^\dag,\hat{\rho}\rbrace \right)
\\
+&(\bar n(\omega_{q}) +1)\left( \hat{a}_{n,q}~\hat{\rho}~\hat{a}_{n,q}^\dag -\frac{1}{2}\lbrace \hat{a}_{n,q}^\dag \hat{a}_{n,q},\hat{\rho}\rbrace \right) \bigg],\nonumber
\end{align}  
where $\gamma_{q}$ and $\bar{n}(\omega_q)=\left(\exp(\beta_q\omega_q)-1\right)^{-1}$ {with }the inverse temperature $\beta_q=(k_B T_q)^{-1}$ of the thermal environment determine the relaxation rates of the oscillators. We note that the bath correlation function (BCF) induced by the Lindblad oscillators, governing electronic dynamics, is given by  \cite{Lemmer2018}
\begin{align}
\label{eq:BCFLindlbad}
C(t)=&\sum_{q,\pm} \bigg[ \frac{\omega_q^2 s_q}{2} \left(\coth\left(\frac{\beta_q \omega_q}{2}\right) \pm 1\right) e^{ \mp  i\omega_q t -\gamma_q t}\bigg].
\end{align}
{The correspondence between the unitary system-environment evolution and our discrete-mode set of Lindblad oscillators underlies the equivalence between the BCFs \cite{Tamascelli2018a}}. This implies that for a given experimentally measured or theoretically computed spectral density, one can fit the corresponding BCF, {at any temperature}, by using Eq.\eqref{eq:BCFLindlbad} for non-perturbative simulations of electronic dynamics. In the supplemental material (SM), an example spectral density is considered, which is benchmarked by analytically solvable models and {very costly} numerically exact HEOM simulations. For simplicity, we consider identical sets of oscillators, hence identical environments, coupled to each site, but our approach can be generalized to asymmetric cases.

\textit{MPO representation}---We now discuss how mixed vibronic states can be described efficiently by MPOs~\cite{Schollwock2011}. A vibronic state can be formally expressed as
\begin{equation}
\label{eq:rho}
\hat{\rho} = \sum_{m,n=1}^N \ketbra{m}{n} \otimes \hat{\mathcal{O}}_{m,n},
\end{equation}
where $\hat{\mathcal{O}}_{m,n}$ describes the vibrational degrees of freedom conditional to the electronic operator $\ketbra{m}{n}$
\begin{equation}
\label{eq:Onm}
\hat{\mathcal{O}}_{m,n} = \sum_{j_1,\ldots,j_M=1}^{N_b^2} C_{j_1,\ldots,j_M}^{(m,n)} \hat x_{j_1} \otimes \ldots \otimes \hat x_{j_M},
\end{equation}
where $N_b$ is the truncation number of the oscillator levels,  \textit{i.e.} the local dimension, and $\{\hat x_i\}_{i=1}^{N_b^2}$ denotes an operator basis that is orthonormal with respect to the Hilbert-Schmidt inner product. A full description of Eq.\eqref{eq:Onm} involves an exponentially large number of coefficients, $N_{b}^{2M}$. We employ MPOs with a bond dimension $\chi$ to describe the operators $\hat{\mathcal{O}}_{m,n}$ in a memory-efficient way. In detail, we encode the coefficients $C_{j_1,\ldots,j_M}^{(m,n)}$ by a set of matrices as
\begin{equation}
\label{eq:OnmMPO}
C_{j_1,\ldots,j_M}^{(m,n)} =  A^{(m,n)}_{1,j_1} A^{(m,n)}_{2,j_2} \cdots A^{(m,n)}_{M,j_M},
\end{equation}
where $A_{i,j_i}^{(m,n)}$ is a matrix of size $d_{i} \times d_{i+1}$ with $d_1 = d_{M+1} = 1$ and $d_i \le \chi$ for $i = 1,\cdots,M+1$ (Fig.\ref{fig:model_MPOrepresentation}(b)). $A_{i,j_i}^{(m,n)}$ is associated with oscillator $i$ and its correlations with the other modes {while $j_i \in \{1,...,N_b^2\}$}. We note that the accuracy of the MPO description is determined by $\chi$. {Typically $\chi$ needs to grow with the {strength} of correlations between oscillators and as long as these correlations are small we can find very good approximations to the full quantum state using a small $\chi$} . For $\chi=1$, only { product states} can be represented, where all the modes are uncorrelated. For sufficiently large $\chi$, any vibronic state can be represented in an exact manner. In summary, Eq.\eqref{eq:OnmMPO} shows that the MPO representation of each $\hat{\mathcal{O}}_{m,n}$ requires $MN_b^2$ matrices with at most $\chi^2$ elements (Fig.\ref{fig:model_MPOrepresentation}(b)).\\

\textit{Dynamics}---The dynamics of $\hat \rho$ in Eq.\eqref{eq:rho} within our MPO description is carried out by the Suzuki-Trotter decomposition of the time-evolution operator $\mathcal{U}(t+\Delta t,t) = \mathcal{U}_e\,\mathcal{U}_L\,\mathcal{U}_v\,\mathcal{U}_{ev} + O(\Delta t^2)$ with $\mathcal{U}_x \equiv \mathcal{U}_x(t+\Delta t,t) = e^{-i \Delta t \hat H_x}$, for $x \in \{e,v,ev\}$, describing Hamiltonian dynamics, and $\mathcal{U}_L$ is responsible for the Lindblad damping. {In principle, other decompositions of the propagator with better error scaling are possible \cite{Hatano2005}.} The action of $\mathcal{U}_v$, $\mathcal{U}_{ev}$, $\mathcal{U}_L$ on $\hat \rho$ can be described by local transformations for each oscillator $i$
\begin{equation}
\label{eq:localupdatesA}
A_{i,j}^{(m,n)}(t+\Delta t) = \sum_{k=1}^{N_b^2} {W}_{x;i,j,k}^{(m,n)} A_{i,k}^{(m,n)}(t),
\end{equation}
where for a given $i$, the evolution of $A_{i,j}^{(m,n)}$ depends only on $A_{i,k}^{(m,n)}$, and the MPO representation of the updated { matrices} preserves the bond dimension. This originates from the fact that $\mathcal{U}_{x=v,ev,L}$ acts locally on each oscillator and hence cannot create correlations among them. The coefficients ${W}_{x;i,j,k}^{(m,n)}$ are determined by the action of $\mathcal{U}_{x=v,ev,L}$ on the oscillator basis $\{\hat x_i\}_{i=1}^{N_b^2}$. {On the other hand,} the action of $\mathcal{U}_{e}$ involves non-local transformations and creates correlations between different oscillators:
\begin{equation}
\label{eq:electronic_update}
\mathcal{U}_{e} \hat{\rho} ~\mathcal{U}_{e}^\dag = \sum_{m,n=1}^N \ketbra{m}{n} \otimes \sum_{k,l=1}^N
 \left[\mathcal{U}_{e}\right]_{m,k} {\left[\mathcal{U}_{e}\right]}_{n,l}^* \hat{\mathcal{O}}_{k,l},
\end{equation}
with $\left[\mathcal{U}_{e}\right]_{m,k} = \bra{m}\mathcal{U}_{e}\ket{k}$ in the site basis $\{\ket{n}\}_{n=1}^N$.  The summation of two MPOs $\hat{O}_{k,l}$, $\hat{O}_{k^\prime,l^\prime}$ does not preserve the bond dimension since it requires the direct sum of their A-matrices $A^{(k,l)}_{i,j_i} \oplus A^{(k^\prime,l^\prime)}_{i,j_i}$ \cite{Hubig2017,Schollwock2011} (Fig.\ref{fig:model_MPOrepresentation}(c)). A compression algorithm (such as the singular value decomposition (SVD)) is required in order to keep the bond dimension of the resulting MPO fixed despite the mixing action of $\mathcal{U}_e$. When an operator $\hat{\mathcal{O}}$ is compressed into $\hat{\mathcal{O}}_{\rm{compr.}}$ with a bond dimension $\chi$, the error $\xi$ can be calculated by {summing} the discarded singular values at every bond of the MPO. The total error $\mathcal{E}$ at every timestep is the sum of all the individual compression errors $\xi$ (see SM). Importantly, the Lindblad damping of Eq.(\ref{eq:damping}) destroys correlations {between modes and effectively reduces the required bond dimension for given desired precision (Fig.\ref{fig:model_MPOrepresentation}(a),(c)). In this work, we employ a time-independent bond dimension $\chi$, although adaptative approaches with high and low values of $\chi$ during the transient and decay of correlations respectively, can optimize the performance significantly. All numerical examples shown in this work were computed with a time step of $\Delta t = 0.5$ fs, which was found to be sufficient to neglect the Trotter error {for the total integration time} up to 1 ps.  

\textit{Scaling of computational resources}---The structure of the DAMPF algorithm is highly parallelizable. Firstly, the local updates in Eq.\eqref{eq:localupdatesA} of the $A^{(m,n)}_{i,j_i}$ matrices are independent from each other and can be carried out in parallel. More importantly, the most time-consuming part of the algorithm, namely, the update of the MPOs after the action of $\mathcal{U}_e$ (Eq.\eqref{eq:electronic_update}), can be parallelized. The compression steps {for each sum related to $\ket{m}\bra{n}$} are independent of each other and can be carried out in parallel. The dependence of the computation time on the system size is polynomial $O( N^5)$ if parallel computing is not exploited (this scaling can be lowered around $O(N^{3.5})$ {by discarding the terms in Eq.(\ref{eq:electronic_update}) with coefficients $\bra{m}\mathcal{U}_{e}\ket{k}$ that are sufficiently small}, although this will depend on the structure and maximum coupling strength $J_{m,n}$ of the network).  We have been able to reduce the scaling to between $O(N^3)$ and $O(N^{2.5})$ with a parallel implementation of the algorithm with 2 processors of 8 cores each. {Further parallelization of the {summation} could, in principle, reduce the scaling to $O(N\log N)$}. The scaling with $N_b$ { has been determined to be approximately} {$O(N_b^2)$} {(for $Q=1$)} and {for the bond dimension} between $O(\chi^{1.5})$, $O(\chi^{2})$, although they slightly depend on the precise implementation of the algorithm. These scaling factors have been numerically determined {and recent techniques could improved them further \cite{Tamascelli2015,Kohn2018}}. 
\begin{figure}
 \centering
	\includegraphics[width=0.49\textwidth]{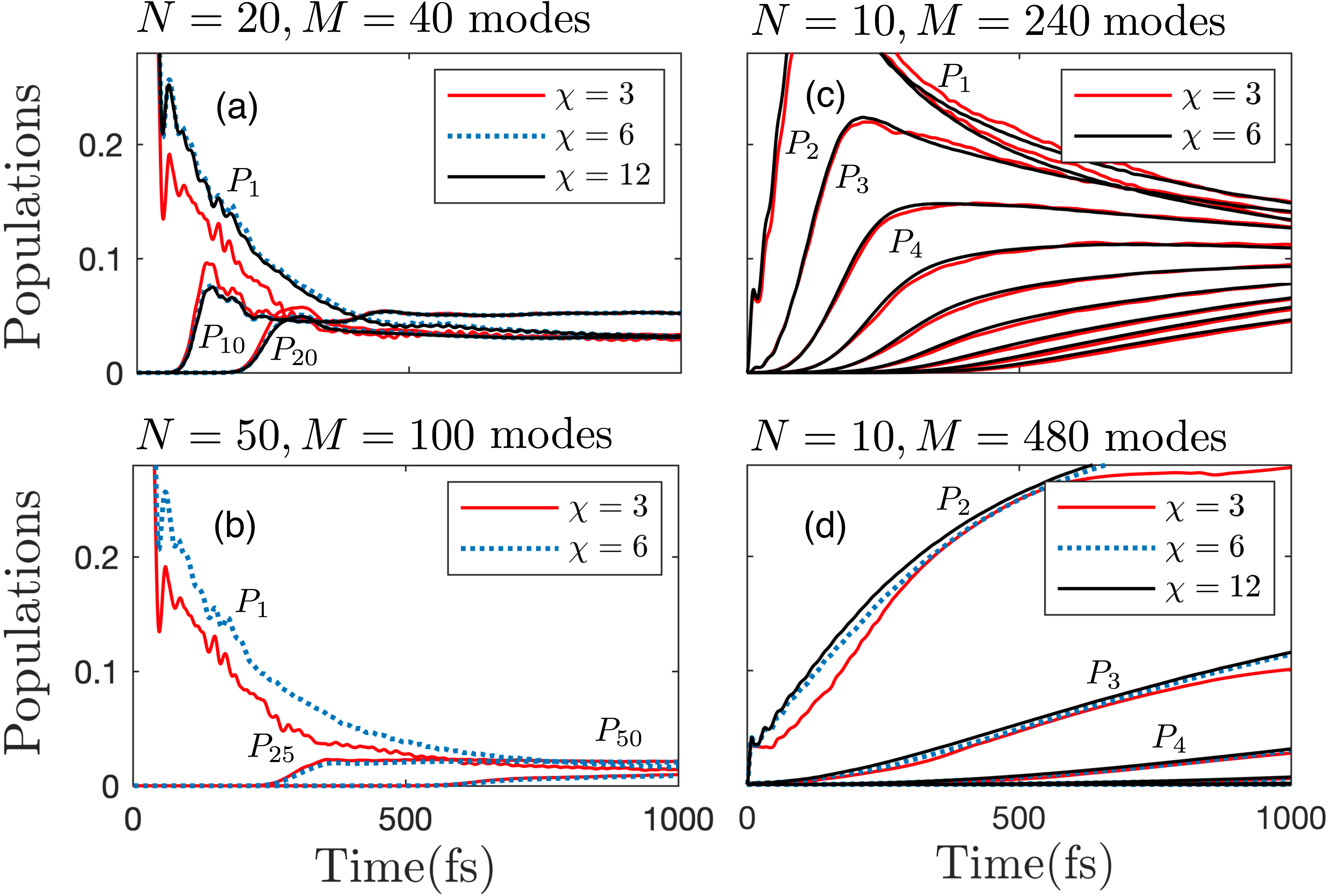} 
\caption{Time evolution of populations $P_k$ of site $k$ in a vibronic chain consisting of (a) $N = 20$ and (b) $40$ sites with $\Omega_m=\Omega_n$, $J_{n,n+1} = 400$ cm$^{-1}$, and two oscillators coupled to each site:  $\omega_1= 1500$ cm$^{-1}$, $s_1=0.5$, $\gamma_1=(1$ ps$)^{-1}$, $T_1=300$ K, $N_{b,1}=8$ and $\omega_2=500$ cm$^{-1}$, $s_2=0.1$, $\gamma_2=(50$ fs$)^{-1}$, $T_2=300$ K, $N_{b,2}=4$. (c) and (d) show the population dynamics of a composite system of $N=10$ sites with (c) $24$ and (d) $48$ modes coupled to each site and $\Omega_m = \Omega_n$, $J_{n,n+1} = 200$ cm$^{-1}$. The vibrational frequencies are uniformly distributed between 400 cm$^{-1}$ and 1500 cm$^{-1}$, with uniform $s_q=0.05$ and $\gamma_q=(200$ fs$)^{-1}$, except for the $400$ cm$^{-1}$ mode that is overdamped by $\gamma^\prime=(50$ fs$)^{-1}$ wtih $s^\prime=0.1$. All the modes are at room temperature $T_q = 300$ K and simulated with $N_b = 5$. In all the simulations, an electronic excitation is initially localized at site 1. Converged results correspond to solid black curves.}	\label{fig:letter_results}
\end{figure}


\textit{Results}---{So far we have discussed how the vibronic dynamics in a spatially extended system can be computed in a memory-efficient way. Here, as an example, we will consider a phenomenological model where an overdamped oscillator is considered as a source of electronic dephasing due to a non-Markovian environment in combination with high-frequency underdamped oscillators that induce coherent vibronic energy transfer. 

In this work, we consider two different vibrational environments. {In Fig.\ref{fig:letter_results}(a) and (b), we first consider the case that each site is strongly coupled to an underdamped mode with frequency 1500 cm$^{-1}$, Huang-Rhys factor {$s_1=0.5$, and $\gamma_1$ = $(1$ ps$)^{-1}$}. Such a large Huang-Rhys factor can be found in polymer-based materials in organic photovoltaics \cite{Coropceanu2002a,Spano2005,Clark2007,Tamura2012a,DeSio2017,Polkehn2018a} and photosynthetic complexes containing carotenoids \cite{Thyrhaug2018}, where vibrational progressions are well pronounced in optical responses. In addition to the underdamped modes, we couple an overdamped mode to each site, and all the modes are coupled to thermal reservoirs at room temperature. Such a two-mode approach has been widely employed in the study of electron transfer \cite{Onuchic1987}.}  In Fig.\ref{fig:letter_results}(a), the energy transfer {through} a linear chain consisting of $N=20$ sites is shown with inter-site couplings  $J_{n,n+1}=400 \,$ cm$^{-1}$ { as an example, although arbitrary network structures can be considered in our approach.} Convergence was achieved for $\chi = 12$ and the corresponding error is $\mathcal{E}\lesssim 10^{-4}$ at every timestep. The simulation time {for the converged solution was 60 hours in a single node of 2 processors with 8 cores each}. {Surprisingly, it is found that the approximate results} with $\chi=6$ are not only qualitatively, but also quantitatively {very close} to the converged results with $\chi=12$. {Even $\chi=3$ already delivers results that exhibit qualitatively the correct dynamics and can be computed in less than 7 hours.} {This implies that one can perform approximate simulations for large systems with low bond dimensions}. For example, as shown in Fig.\ref{fig:letter_results}(b), we can simulate a vibronic chain consisting of $N=50$ sites in an approximate manner with a bond dimension of $\chi=6$ {(12 days of simulation time)}. We note that large Huang-Rhys factors for such system sizes represent a major challenge for other methods. For example, the memory cost of HEOM quickly exceeds one terabyte when the system size $N>10$ (see SM), {limiting simulations drastically both in memory and computation time}. In contrast, the memory requirement of our method for $N = 20$ sites, shown in Fig.\ref{fig:letter_results}(a), is less than 300 megabytes for $\chi=6$ and around 1.2 gigabytes for the converged solution with $\chi=12$. We also notice that $\chi=6$ provides converged results when $\gamma_1$ {is increased to $(500$ fs$)^{-1}$}, {demonstrating that the Lindblad damping reduces the required bond dimension for {numerically} exact simulations.} 

Secondly, in Fig.\ref{fig:letter_results}(c) and (d) {we consider} a different vibrational structure where each site is coupled to multiple high-frequency modes with relatively small Huang-Rhys factor of 0.05 for each mode and $J_{n,n+1}=200 \,$ cm$^{-1}$. Such a rich vibrational spectrum is a typical feature of photosynthetic pigment-protein complexes \cite{Blankenship2002book}. In Fig.\ref{fig:letter_results}(c) and (d), we show the energy transfer through a linear chain consisting of $N=10$ sites where each site is locally coupled to $Q=24$ and $48$ modes, respectively. The vibrational frequencies are uniformly distributed between 400 cm$^{-1}$ and 1500 cm$^{-1}$, with uniform $s_q=0.05$ and $\gamma_q=(200$ fs$)^{-1}$, except for the $400 $ cm$^{-1}$ mode that is overdamped to induce electronic dephasing. Converged results are obtained for bond dimensions close to (a) $\chi=6$ and (b) $\chi=12$, while $\chi=6$ already provides a good approximation. As the coupling strength to the vibrational environment $\lambda=\sum_q \omega_q s_q$ is doubled {(from $\lambda=1160$ cm$^{-1}$ in Fig.\ref{fig:letter_results}(c) to $\lambda=2320$ cm$^{-1}$ in Fig.\ref{fig:letter_results}(d))}, polaron formation slows down the transport of the electronic excitation. Increasing the number of local modes to $Q=100$ {($M=1000$ modes in total)} completely localizes the excitation in the first site (see SM). The simulation times for $\chi=6$ were 10, 19, 33 hours for $Q=24$, 48, 100 modes, respectively. Due to the relatively low local dimension $N_b = 5$ required for exact simulations, the inclusion of additional modes does not {impose} a {significant} numerical cost due to the {more} efficient MPO representation. {The case with $\lambda \sim J_{n,n+1}$ also can be efficiently simulated by our approach (see SM)}.

\textit{Conclusions}---These results show that our proposed method is able to simulate with moderate simulation cost the dynamics of spatially extended vibronic systems with highly structured environments under a wide range of parameter regimes: from coherent vibronic dynamics to dynamic localization with polaron formation. It is notable that numerically exact simulations with $N\geq 10 $ sites and $Q\geq 100$ local modes and moderate Huang-Rhys factors are possible with DAMPF,  which brings the actual dynamics of interesting natural molecular aggregates \cite{Valkunas_book,Scholes2017,Wendling2000a,Novoderezhkin2005} within the realm of exact simulations. This will facilitate the investigation of the role of highly structured vibrational environments of photosynthetic complexes \cite{Huelga2013,Chin2010NJP,Chin2013,Womick2011JPCB,Chin2012PhilTransA,Kolli2012JCP,Tiwari2013PNAS,Christensson2012JPCB,Irish2014PRA}, which have been severely coarse-grained in previous theoretical studies due to the limited computational power \cite{Blau2017a}. In addition, DAMPF can be used to compute spectroscopic quantities, such as absorption spectra (see SM), suggesting the possibility to study the connection between spectroscopic observations and underlying energy transfer dynamics, therefore assisting in the identification of physical mechanisms that may lead to diverse observations \cite{Panitchayangkoon2010,Thyrhaug2018a,Blau2017a,Miller2017PNAS,Scholes2018NatChem,Collini2010,ScholesJPCB2018}. Finally, the non-equilibrium dynamics of underdamped modes can also be monitored, making our approach very suitable for the study of the role of strongly coupled vibrational modes in organic photovoltaics where coherent vibronic coupling has been suggested to promote charge separation \cite{Falke2014,DeSio2016}.} 
 
\textit{Acknowledgments}---We thank Andrea Smirne and Andreas Lemmer for helpful discussions. This work was supported by the ERC Synergy grant BioQ and the Templeton Foundation.

\bibliographystyle{/usr/local/texlive/2016/texmf-dist/bibtex/bst/revtex/apsrev4-1}

\begin{widetext}
\section{Supplemental Material}
\subsection{Model}
The vibronic system is modeled by a network of $N$ electronically coupled sites, where each site interacts with local oscillators connected to a Markovian bath (see Fig.1(a) in the main manuscript). We assume that every site is connected to the same number of oscillators, although arbitrary sets of oscillators per site can be taken into account in our approach. Accordingly, each site is locally coupled to $Q$ harmonic oscillators and the total number of oscillators is denoted by $M=NQ$. We will restrain our method to the single-excitation manifold, with $\ket{n}$ representing an excitation at site $n$. This is sufficient to describe linear optical responses, such as absorption by including the global electronic ground state in the model, and to investigate energy transfer dynamics under low light conditions. The vibronic system is described by the following Hamiltonian $\hat{H}_{\rm S}$
\begin{align}
\label{eq:SMhamiltonian}
\hat H_{\rm S} &= \hat H_e + \hat H_v + \hat H_{ev},\\
\label{eq:SMHe}
\hat H_e &= \sum_{n=1}^N \Omega_n \ketbra{n}{n} + \sum_{m\neq n} J_{m,n} \ketbra{m}{n}, \\
\label{eq:SMHv}
\hat H_v &= \sum_{n=1}^N  \sum_{q=1}^Q \omega_{q} \hat a_{n,q}^\dag \hat a_{n,q}, \\
\label{eq:SMHev}
\hat H_{ev} &= \sum_{n=1}^N\sum_{q=1}^Q \omega_{q} \sqrt{s_{q}} \ketbra{n}{n}  \left(\hat a_{n,q} + \hat a_
 {n,q}^\dag \right),
\end{align}
\noindent where $\Omega_n$ is the excitation energy of the $n$-th monomer, $J_{m,n}$ is the electronic coupling between sites $n$ and $m$, $\hat{a}_{n,q}$ and $\hat{a}_{n,q}^\dag$ denote the annihilation and creation operators, respectively, of the $q$-th oscillator locally coupled to site $n$, with vibrational frequency $\omega_q$ and Huang-Rhys factor $s_q$.

In order to faithfully retain the non-Markovian character of the non-equilibrium dynamics of the vibronic system, we split the system into a non-Markovian core that captures the vibronic features which, in turn, is coupled to a Markovian environment, 
\begin{align}
\label{eq:SMHbath}
\hat H_{\rm Bath}&=\sum_{n=1}^N  \sum_{q=1}^Q \sum_{\xi} \omega_\xi \hat{b}_{n,q,\xi}^\dag \hat{b}_{n,q,\xi}, \\
\label{eq:SMHvbath}
\hat H_{v\rm-Bath} &=\sum_{n=1}^N  \sum_{q=1}^Q  \sum_{\xi} g_{q,\xi} \left(\hat{b}_{n,q,\xi}^\dag + \hat{b}_{\xi,n,q}\right) \left(\hat{a}_{n,q} + \hat a_{n,q}^\dag \right),
\end{align}
\noindent where  Eq.(\ref{eq:SMHbath}) is the bath Hamiltonian, and Eq.(\ref{eq:SMHvbath}) describes the coupling between oscillators $\hat a_{n,q}$ and bath modes $\hat b_{n,q,\xi}$. The dynamics of the vibronic state $\hat\rho$ under the thermal secondary bath is described by a Markovian quantum master equation
\begin{equation}
\label{eq:SMmastereq}
\frac{d}{dt}\hat{\rho}=-i[\hat{H}_{\rm S},\hat \rho] + \mathcal{D}\hat{\rho},
\end{equation}
\noindent where the Lindblad damping of the oscillators is described by the dissipator $\mathcal{D}$ 
\begin{align}
\label{eq:SMDissipators}
\mathcal{D}\hat{\rho}= \sum_{n=1}^N\sum_{q=1}^Q & \bigg[ \gamma_{q} (\bar n(\omega_{q}) +1)\left( \hat{a}_{n,q}~\hat{\rho}~\hat{a}_{n,q}^\dag -\frac{1}{2}\lbrace \hat{a}_{n,q}^\dag \hat{a}_{n,q},\hat{\rho}\rbrace \right)  + {\gamma_{q}} \bar n(\omega_{q}) \left( \hat{a}_{n,q}^\dag~\hat{\rho}~\hat{a}_{n,q} -\frac{1}{2}\lbrace \hat{a}_{n,q} \hat{a}_{n,q}^\dag,\hat{\rho}\rbrace \right)\bigg],
\end{align} 
\noindent characterized by the damping rates $\gamma_{q}$ and the mean phonon number $\bar{n}(\omega_q)=\left(\exp(\beta_q\omega_q)-1\right)^{-1}$  of the oscillator with frequency $\omega_q$ at inverse temperature  $\beta_q=(k_B T_q)^{-1}$.

\subsection{Dynamics}
We approximate the unitary part of the master equation Eq.\eqref{eq:SMmastereq} by the well-known Suzuki-Trotter decomposition of the evolution operator $\mathcal{U}(t+\Delta t,t) = e^{-i \Delta t \hat H}$:
\begin{equation}
\label{eq:SMpropagator}
\mathcal{U}(t+\Delta t,t) = \mathcal{U}_e \mathcal{U}_L \mathcal{U}_v\mathcal{U}_{ev} + o(\Delta t^2),
\end{equation}
with $\mathcal{U}_x \equiv \mathcal{U}_x(t+\Delta t,t) = e^{-i \Delta t \hat H_x}$, for $x \in \{e,v,ev\}$ and $\mathcal{U}_L$ is responsible for the Lindblad damping. The evolution of the state $\hat \rho$ is determined by the action of the three contributions of the propagator Eq.\eqref{eq:SMpropagator} on the MPO representation of $\hat\rho$.

The state evolved under $\mathcal{U}_v$, $\mathcal{U}_L$, and $\mathcal{U}_{ev}$ exhibits a representation with the same bond dimension as the initial state. This originates from the fact that the vibrational and the interaction part act locally on each oscillator and hence cannot create correlations among them. Formally, the state $\mathcal{U}_x \hat\rho_S\mathcal{U}_x^\dag$, for $x \in \{v,ev\}$, exhibits an MPO representation of the form
\begin{equation}
\label{eq:SMtransform A matrix v,ev}
A_{i,j}^{(m,n)} (t+\Delta t)= \sum_{k=1}^{N_b^2} {W}_{x;i,j,k}^{(m,n)} A_{i,k}^{(m,n)}(t),
\end{equation}
 where $\{A_{i,j}^{(m,n)}\}$ is an MPO representation of the initial state $\hat\rho$. In particular, the matrices of the MPO representation associated with the $i$-th oscillator are updated locally. In equation Eq.\eqref{eq:SMtransform A matrix v,ev}, the coefficients are determined by the action of the evolution on the oscillator basis $\{\hat x_i\}$. That is, for the vibrational part $x=v$, we obtain ${W}_{v;i,j,k}^{(m,n)} \equiv {W}_{v;i,j,k}$, with
\begin{equation}
{W}_{v;i,j,k} = \Tr \left( \hat x_j^\dag \mathcal{U}_v^{(n,q)} \hat x_k \mathcal{U}_v^{(n,q)\dag} \right),
\end{equation}
where $\mathcal{U}_v^{(n,q)} = \exp(-i \Delta t \omega_{n,q} \hat a_{n,q}^{\dag}\hat a_{n,q})$ and $i = (q-1)N + n$, for all $n= 1,\ldots,N$ and $q = 1,\ldots,Q$. 
On the other hand, due to the local form of the interaction term Eq.~\eqref{eq:SMHev}, the only nonzero coefficients for $x = ev$ are for terms associated with electronic populations and coherences, respectively, given by
\begin{equation}
{W}_{ev;i,j,k}^{(n,n)} = \Tr \left( \hat x_j^\dag \mathcal{U}_{ev}^{(n,q)} \hat x_k \mathcal{U}_{ev}^{(n,q)\dag} \right),
\end{equation}
if $i = (q-1)N + n$ for $q = 1,\ldots,Q$, and (for $m \ne n$)
\begin{equation}
{W}_{ev;i,j,k}^{(m,n)} = \begin{cases} \Tr \left( \hat x_j^\dag \mathcal{U}_{ev}^{(m,q)} \hat x_k \right), &\text{ if } i = (q-1)N+m, \\ \Tr \left( \hat x_j^\dag \hat x_k \mathcal{U}_{ev}^{(n,q)\dag} \right), &\text{ if } i = (q-1)N+n. \end{cases}
\end{equation}
where $\mathcal{U}_{ev}^{(n,q)} = \exp(-i \Delta t \omega_{n,q} \sqrt{S_{n,q}} (\hat a_{n,q} + \hat a_{n,q}^{\dag}))$.

In contrast, the electronic contribution to the evolution requires the summation of different MPOs. The action of the evolution on the state $\hat \rho$ takes the form 
\begin{equation}
\label{eq:SMUerhoUe}
\mathcal{U}_{e} \hat{\rho}_S \mathcal{U}_{e}^\dag = \sum_{m,n} \ketbra{m}{n} \otimes \sum_{k,l=1}^N
 \left[\mathcal{U}_{e}\right]_{m,k} {\left[\mathcal{U}_{e}\right]}_{n,l}^* \hat{\mathcal{O}}_{k,l},
\end{equation}
with $\left[\mathcal{U}_{e}\right]_{m,n} = \bra{m}\mathcal{U}_{e}\ket{n}$ being the matrix elements of the evolution operator in the electronic basis $\{\ket{n}\}_{n=1}^N$ and $º\alpha^*$ denotes the complex conjugate of $\alpha \in \mathbb{C}$. Since adding MPOs may increase the bond dimension \cite{Schollwock2011}, the evolution under the electronic contribution does not preserve the bond dimension of the oscillator part. In order to control the growth of the dimension of the MPOs we employ the SVD compression scheme \cite{Verstraete2006} that truncates the matrices of the MPO representation dimension. We apply the compression scheme after every sum of two terms, however, other schemes are possible. The error introduced by the compression can be estimated and thus the quality of the approximation can be monitored during the time evolution.

\subsection{Reduced electronic and vibrational states}

In order to efficiently take the partial trace over the oscillator degrees of freedom, we choose the basis $\{\hat x_i\}$ with $\hat x_1 = N_b^{-1/2}\id$ and $\hat x_i$ traceless for $i > 1$. The reduced electronic density matrix of the state $\hat\rho_S$ thus becomes
\begin{equation}
\hat{\rho}_e \equiv \Tr_v\!\left( \hat\rho_S \right) = N_b^{M/2} \sum_{m,n=1}^N \left(\prod_{i=1}^M A_{i,1}^{(m,n)}\right) \ketbra{m}{n},
\end{equation}
with $\Tr_v(\cdot)$ denoting the partial trace with respect to the vibrational degrees of freedom. The reduced density matrix for the vibrational degrees of freedom can be readily calculated,
\begin{align}
\hat \rho_v = \Tr_e \hat \rho = \sum_{n=1}^N \hat{\mathcal{O}}_{n,n}  
 = \sum_{l_1,\ldots,l_{M} = 1}^{N_b^2} B_{1,l_1} \cdots B_{M,l_M} \;\hat x_{l_1} \otimes \cdots \otimes \hat x_{l_{M}},
\end{align}
with $B_{i,l_i} = \oplus_{m_i=1}^{N_b^2} A_{i,m_i}^{(n,n)}$. In order to investigate non-classicality of the molecular vibrations we need to access the density matrix of a particular oscillator $\alpha$,
\begin{align}
\Tr_{\bar{\alpha}} \hat{\rho}_v = N_b^{(M-1)/2} \sum_{l_\alpha} B_{1,1} \cdots B_{\alpha,l_{\alpha}}\cdots B_{M,1} \;  \hat x_{l_\alpha}
\end{align}
\noindent where we used the notation $\Tr_{\bar{\alpha}}$ to indicate the trace operation with respect all oscillators except $\alpha$.

\subsection{Error bound}
In the following we introduce an error bound for the SVD compression scheme that we apply at every time step. 
First, recall that any $M$-body operator $\hat{\mathcal{O}}$ can be approximated by an MPO $\hat{\mathcal{O}}_{\rm{compr.}}$ with a desired bond dimension $\chi$ employing SVD compression as prescribed in \cite{Verstraete2006,Schollwock2011} for matrix product states. The error that is introduced when we compress an operator can be bounded as
\begin{equation}
\label{eq:SMSVD compr error}
\|\hat{\mathcal{O}}-\hat{\mathcal{O}}_{\rm{compr.}} \|_F^2 \le 2 \sum_{i=1}^{M-1} \epsilon^{(i)},
\end{equation}
where $\|\hat{A}\|_F = \Tr(\hat{A}^\dag \hat{A})^{1/2}$ is the Frobenius norm of an operator $\hat A$. Here, $\epsilon^{(i)}$ is the truncation error (the sum of the discarded squared singular values) at bond $i$, i.e.,
\begin{equation}
\epsilon^{(i)} = \sum_{j>\chi} \sigma_{j}\left(C_i\right)^2,
\end{equation}
with $\sigma_i(A)$ denoting the $i$-th largest singular value of $A$ and $C_i$ the $N_b^i \times N_b^{M-i}$ matrix with entries $[C_i]_{(j_1 \ldots j_i)(j_{i+1} \ldots j_M)} = \Tr(\hat{\mathcal{O}}\hat{x}_{j_1} \otimes \cdots \otimes\hat{x}_{j_M})$ for an orthonormal operator basis $\{x_i\}$.

An overall bound of the error on the state collected at every Trotterization step can now be determined. Namely, as noted previously, the evolution under $\mathcal{U}_e$ may increase the bond dimension since updating a given operator $\hat{\mathcal{O}}_{m,n}$ requires a sum over $N^2$ terms, see Eq.\eqref{eq:SMUerhoUe}. Because we choose to apply the SVD compression at every sum of two terms (other schemes are possible) we introduce an error $\xi_{m,n}^{(i)}$ of the form Eq.\eqref{eq:SMSVD compr error} at the $i$-th sum. With this, a bound on the total error incurred at every timestep takes the form
\begin{equation} \label{eq:SMerrorbound}
\mathcal{E}=\|\hat\rho_S - \hat\rho_{S,\rm{compr.}}\|_F^2 \le \sum_{m,n=1}^N \sum_{i = 1}^{N^2-1} \xi_{m,n}^{i},
\end{equation}
where $\hat\rho_{S,\rm{compr.}}$ denotes the state resulting from the above compression procedure.

\clearpage

\subsection{Numerically exact DAMPF simulations based on model spectral density}
In order to demonstrate that one can perform numerically exact simulations of electronic dynamics for a given spectral density $\mathcal{J}(\omega)$, we consider a model spectral density consisting of broad background $\mathcal{J}_b(\omega)$, responsible for electronic dephasing, and a narrow Lorentzian function $\mathcal{J}_u(\omega)$, describing coherent electronic-vibrational interaction
\begin{align}
\label{eq:SMJw}
\mathcal{J}(\omega)=\mathcal{J}_b(\omega)+\mathcal{J}_u(\omega).
\end{align}
The background $\mathcal{J}_b(\omega)$  is modeled by an Ohmic spectral density with a Gaussian cutoff,
\begin{align}
\label{eq:SMgaussianJw}
\mathcal{J}_b(\omega)= \frac{2}{\sqrt{\pi}\Gamma_b \left(1+ \rm erf(\Omega_b/\Gamma_b)\right) } \lambda_b \omega \exp \left( -\left(\frac{\omega-\Omega_b}{\Gamma_b}\right)^2\right),
\end{align}
\noindent which is centered at $\Omega_b \approx 700$ cm$^{-1}$ with a width $\Gamma_b = 500 $ cm$^{-1}$, and the reorganization energy $\lambda_b = \int_{0}^{\infty} d\omega J_b(\omega)/\omega $ is taken to be $500$ cm$^{-1}$. The Lorentzian peak $J_u(w)$ is modeled by
\begin{align}
\label{eq:SMunderdamped}
\mathcal{J}_u(\omega)= \frac{4 \Gamma_u \Omega_u (\Omega_u^2+\Gamma_u^2) s_u}{\pi} \frac{\omega}{\left(\left(\omega+\Omega_u\right)^2 + \Gamma_u^2\right)\left(\left(\omega-\Omega_u\right)^2+\Gamma_u^2\right)}
\end{align}
describing an underdamped mode with vibrational frequency $\Omega_u =1500$ cm$^{-1}$ and a damping rate $\Gamma_u = (500$ fs$)^{-1}$. The reorganization energy contribution is given by $\Omega_u s_u = \int_0^{\infty}d\omega \mathcal{J}_u(\omega)/\omega$. These parameters are chosen to model the spectral density of organic photovoltaic (OPV) materials computed by density function theories \cite{Tamura2012a}. We note that the Gaussian cutoff is considered in order to make the spectral density does not have a long tale at high frequencies beyond 2000 cm$^{-1}$ (see Fig.\ref{fig:SMBFC}(a)). The Huang-Rhys factor $s_u=0.1$ of the underdamped mode is relatively smaller than the typical values of the C=C stretch modes of the OPV materials, which is of the order of 1.0. We note that such a large Huang-Rhys factor can be considered in our simulations, and these results will be presented elsewhere.\\

Fig.\ref{fig:SMBFC}(b) shows the bath correlation function (BCF) of the model spectral density, defined by
\begin{align}
\label{SMBCF}
C(t) = \int_0^\infty d\omega \, \mathcal{J}(\omega) \left(\coth\left(\frac{\beta\omega}{2}\right) \cos(\omega t) - i \sin(\omega t)\right)
\end{align}
where the temperature of the harmonic bath is taken to be $T=(k_B \beta)^{-1}=300$ K.  We note that the reduced electronic dynamics, where the time evolution of the total system-environment state is described by a global unitary operator, can be equivalent to that under the discrete oscillator environment with Lindblad damping that we consider in our DAMPF approach. The equivalence can be achieved by fitting the effective BCF under the discrete oscillator environment (see Eq.(3) in the main manuscript) to the target BCF in Eq.\eqref{SMBCF}. We note that the equivalence is based on the assumption that the electronic and vibrational states are separable at the initial time (no initial correlations). {For the case of the target environment modelled by the spectral density $\mathcal{J}(\omega)$, the environment is initially in a thermal state at the temperature $T = (k_B \beta)^{-1}$ (see the target BCF in Eq.\eqref{SMBCF}), while in the Lindblad simulations, each oscillator ·$\hat{a}_{n,q}$ is initially in a thermal state at temperature $T_q$ (no initial correlations amongst modes)}. As an example, Fig.\ref{fig:SMBFC}(b) shows that the target BCF can be fitted by 21 Lindblad oscillators, where 20 modes are considered to fit the BCF induced by background $\mathcal{J}_b(\omega)$, while a single mode is employed to fit the BCF by a narrow Lorentzian $\mathcal{J}_u(\omega)$. We note that the Lorentzian peak not only induces underdamped oscillations in the BCF, but also additional exponentially decaying terms (Matsubara terms) with relatively small amplitudes. The parameters of the Lindblad oscillators are numerically optimized for the target BCF (see Table \ref{table:SMoscillator_values}). We note that the target BCF can be fitted with smaller or larger number of Lindblad oscillators, with the fitting quality maintained (not shown here). Larger number of Lindblad oscillators does not necessarily make the DAMPF simulations more costly, because when more modes are considered in the fitting, each mode can have a smaller Huang-Rhys factor $s_q$. This enables one to consider lower local dimensions, which can make the MPO-based DAMPF simulations efficient.\\

\begin{figure}
	\includegraphics[width=0.6\textwidth]{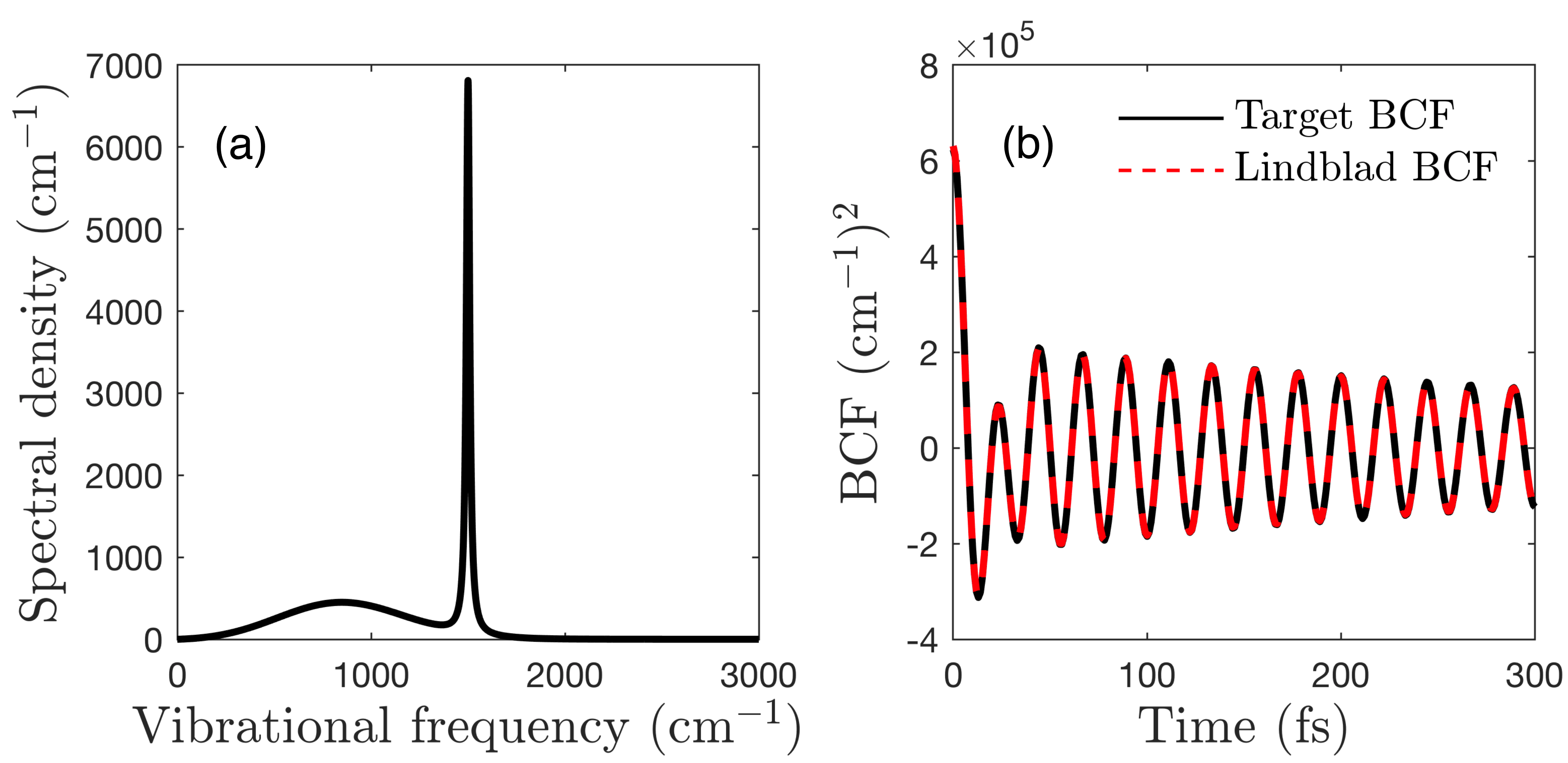} 
	\caption{(a) Model spectral density consisting of broad background centered at around 700 cm$^{-1}$, and a narrow Lorentzian peak centered at 1500 cm$^{-1}$. (b) Corresponding bath correlation function (BCF). The target BCF computed based on the model spectral density, shown in black, is fitted by using the effective BCF of Lindblad oscillators (see Eq.(3) in the main manuscript), shown as a dashed red line.}\label{fig:SMBFC}
\end{figure}
\begin{figure}
	\includegraphics[width=0.7\textwidth]{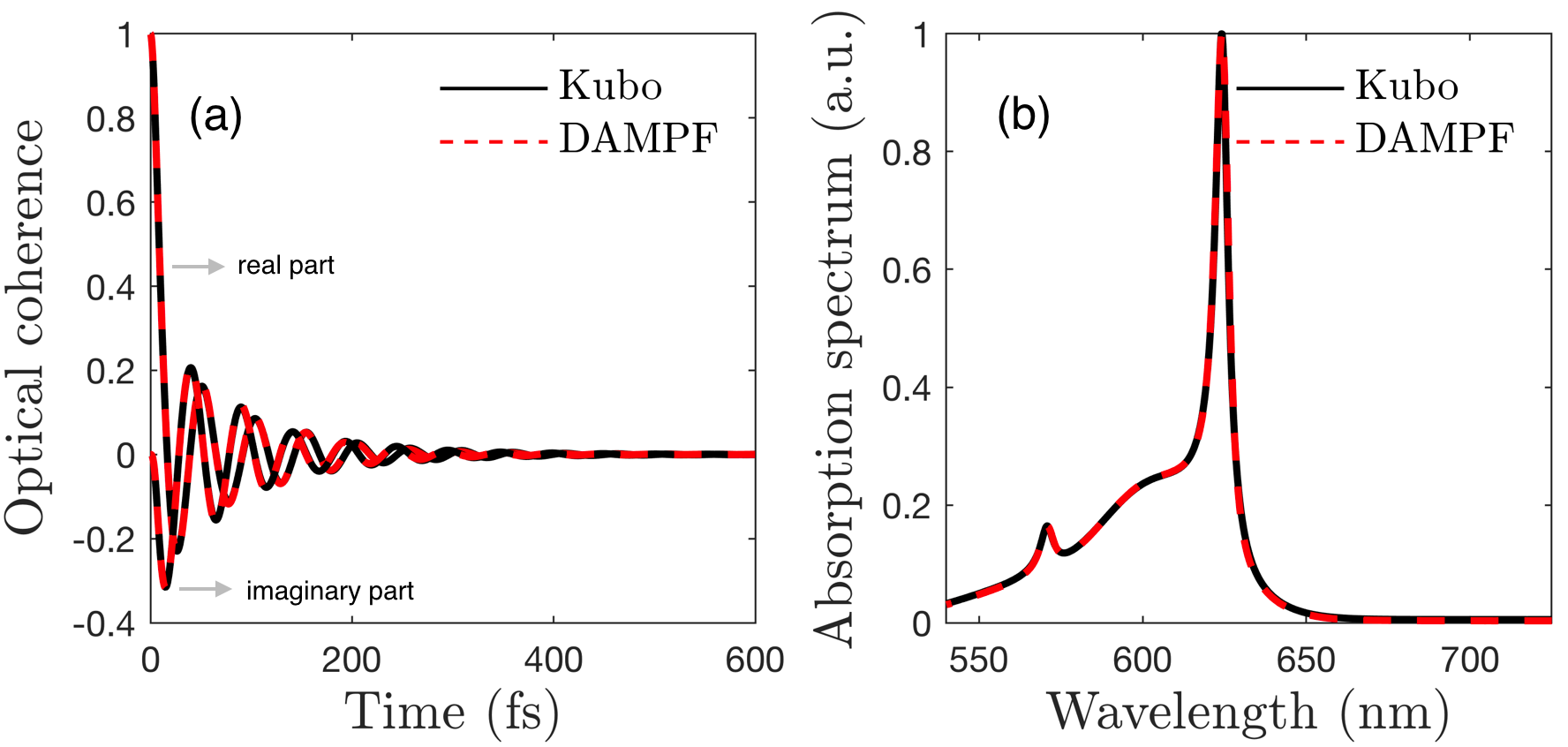} 
	\caption{(a) Optical coherence between electronic ground and excited states of a two-level system (monomer), which can be computed analytically, as shown in black. Numerical results based on DAMPF, shown in red, are well matched to the analytical solution. (b) Absorption lineshape obtained by Fourier transforming the optical coherence dynamics. Absorption peak is red-shifted from the energy-level $\Omega_1$ of the monomer, which is taken to be 600 nm. This is due to the reorganization process of the vibrational environment. In (a), the phase evolution at optical frequency, namely {$e^{i\Omega_1 t}$}, is removed for better visibility, where the remaining slow features are responsible for the red-shift of the absorption peak.} 	\label{fig:SMabsorption}
\end{figure}
\begin{figure}
	\includegraphics[width=0.4\textwidth]{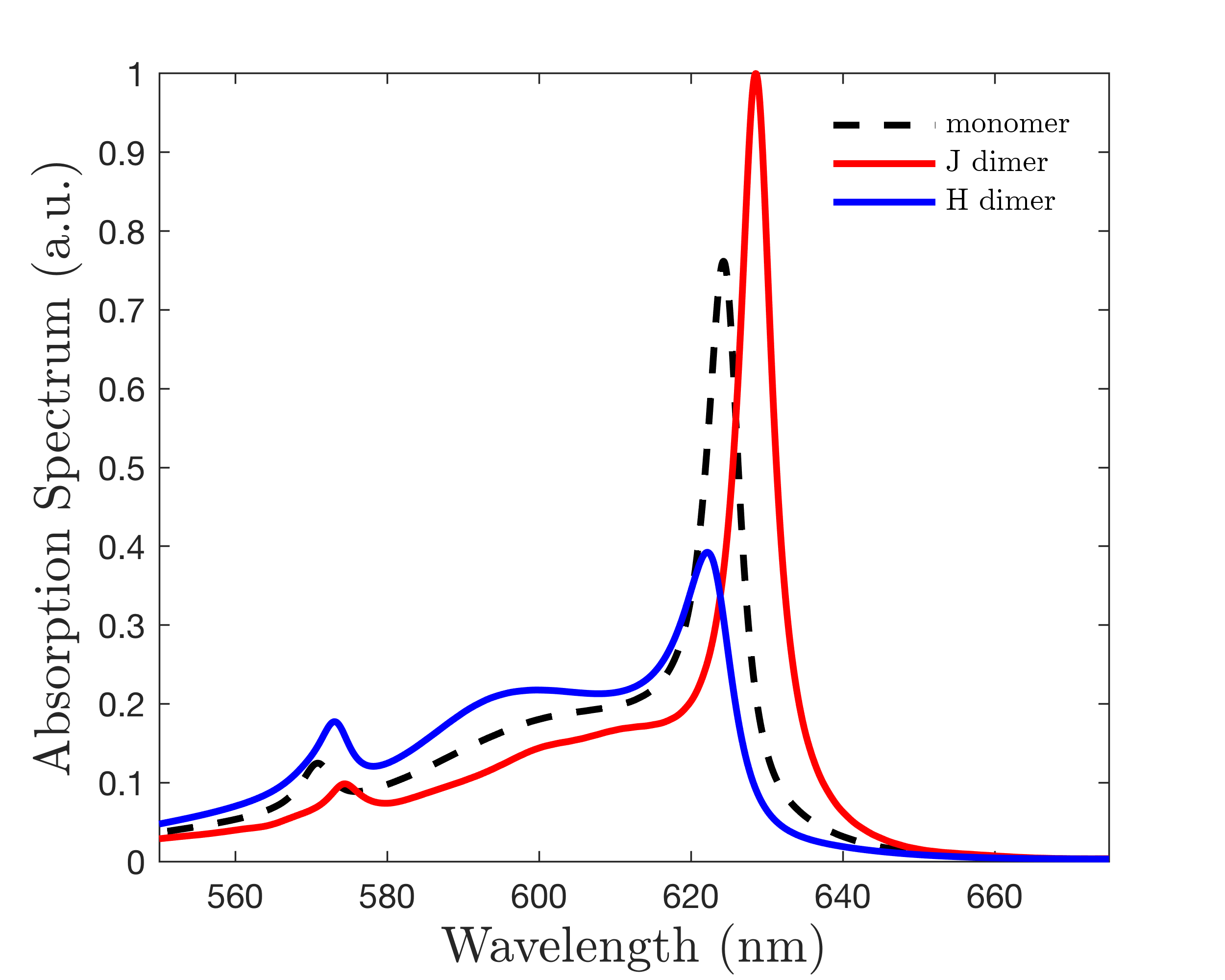} 
	\caption{Absorption spectra of H- and J-dimers, where both sites have the same energy-level, 600 nm, and inter-site coupling is taken to be $J_{12} = \pm 200$ cm$^{-1}$. It is assumed that the two sites have the same transition dipole moment in amplitude and orientation.} 	\label{fig:SMaggabs}
\end{figure}
\begin{table} 
\caption{Summary of the parameters of Lindblad oscillators considered in Fig.(\ref{fig:SMBFC})-(\ref{fig:SMaggabs}). The 21-th oscillator describes the long-lived oscillations in the BCF function induced by the narrow feature of the spectral density (see Fig.\ref{fig:SMBFC} (a)).}\label{table:SMoscillator_values}
\begin{ruledtabular}
\begin{tabular}{ccccc}	
$q$ & $\omega_q$ (cm$^{-1})$  & $\gamma_q^{-1}$ (fs) & $s_q$   & $T_q (K)$\\
\hline
1&        189.94      &       73.28    &   0.019021    &      273.29 \\
2&        243.70      &       64.88      &   0.056442    &      350.63\\
3&       319.72       &       59.69     &   0.072136    &     332.23\\
4&       393.53       &       62.67     &   0.081439    &     294.48\\
5&       464.80       &      65.39      &   0.086793    &     279.72\\
6&       538.00       &      71.49      &   0.087207    &     273.12\\
7&       611.12       &      80.72     &   0.083937    &     266.90\\
8&       687.36       &      93.44   &   0.076730    &     259.44\\
9&       767.21       &    105.06  &   0.066693    &     246.55\\
10&     848.34       &    116.53  &   0.055568    &     192.57\\
11&     928.56       &    126.93   &   0.044172    &     156.98\\
12&   1008.80       &    132.49  &   0.033551    &     116.13\\
13&   1090.30       &    137.00  &   0.024248    &     122.08\\
14&   1172.00       &    143.92  &   0.016661    &     125.21\\
15&   1253.10       &    145.91 &   0.010841    &     156.13\\
16&   1335.10       &    138.63 &   0.006618    &     172.19\\
17&   1417.60       &    136.75 &   0.003757    &     156.87\\
18&   1501.60       &    128.09 &   0.001941    &     184.67\\
19&   1564.00       &    179.54   &   0.000903    &       238.72\\
20&   1599.50       &    72.700    &   0.000009    &         0.000\\
21&   1500.00       &    500.00     &   0.099133    &     130.28\\
\end{tabular}
\end{ruledtabular}
\end{table}

To demonstrate the accuracy of DAMPF results, we compute the absorption spectrum of a two-level system, which is analytically solvable \cite{Kubo2006,BreuerBook}. The absorption lineshape is computed by Fourier transforming the dynamics of the optical coherence between electronic ground and excited states \cite{MukamelBook}, where the vibrational environment is initially in a thermal equilibrium state at temperature $T = 300$ K. It is notable that DAMPF simulated results are quantitatively well matched the analytical results, as shown in Fig.\ref{fig:SMabsorption} (a) and (b), displaying optical coherence dynamics and absorption spectrum, respectively. The DAMPF method can be employed to compute absorption spectra of multi-site systems. As an example, in Fig.(\ref{fig:SMaggabs}), we consider H- and J-dimers, where the transition dipole moments of two sites are in parallel, but the electronic coupling between them is positive- and negative-valued, respectively, with $J_{12}=\pm \,200 $ cm$^{-1}$. For the J-dimer, it is found that absorption peak location is red-shifted, and phonon sideband is suppressed compared to the monomer case. On the other hand, for the H-dimer, absorption peak location is blue-shifted and phonon sideband is enhanced. For all the cases, the absorption cross-sections are normalized. These results are in line with the optical properties of J- and H-aggregates  \cite{Hestand2018}.

\subsection{Some further results}
In Fig.2(c) and (d) in the main manuscript, we consider the case that each site is locally coupled to $Q=24$ and $48$ oscillators, respectively, with the uniform Huang-Rhys factor of $s_q = 0.05$ (see the main manuscript for further detail). It is shown that as the number of local oscillators increases, the electronic excitation transfer is suppressed due to the polaron formation. In Fig.\ref{fig:SMfurtherres}(a), it is demonstrated that when each site is coupled to $Q=100$ oscillators, the electronic excitation stays localized at the initially populated site 1 with a high probability. This is due to the stronger overall environmental coupling strength, $\lambda = \sum_q  \omega_q s_q = 4750$ cm-1, than the cases of $Q=24$ and 48 considered in the main manuscript.

In the above examples, we investigated the cases that the overall environmental coupling is stronger than the inter-site system couplings. To demonstrate that our method can simulate the most non-perturbative regime, in Fig.\ref{fig:SMfurtherres}(b), the overall Huang-Rhys factors are reduced in such a way that the overall environmental coupling strength is the same to inter-site coupling strength, $\lambda = J_{n,n+1} = 200$ cm$^{-1}$. Here we consider a vibronic chain consisting of 10 sites where each site is coupled to $Q=48$ oscillators whose damping rates are even decreased to $\gamma_q = (1 $ps$)^{-1}$. Importantly, it is found that numerically exact results can be obtained with moderate simulation cost (local dimension $N_b$ = 5 and bond dimension $\chi=12$).

\begin{figure}
	\includegraphics[width=0.8\textwidth]{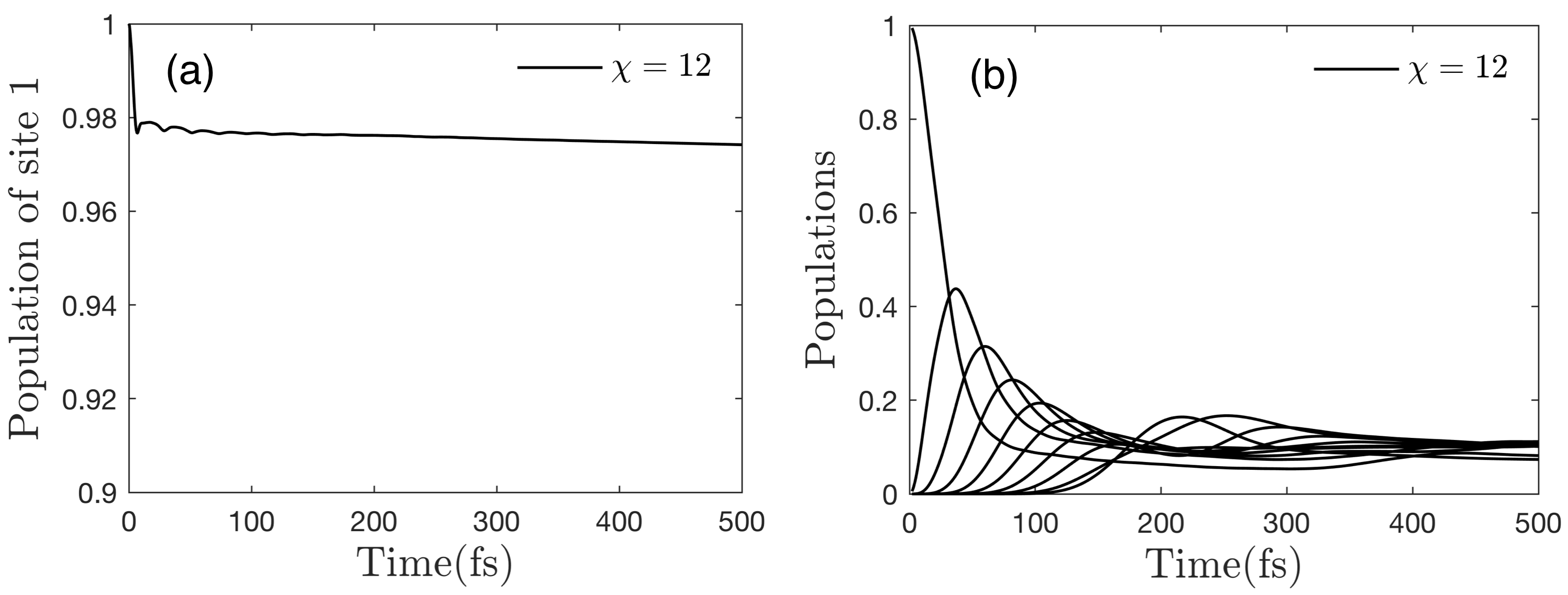}
\caption{(a) Population dynamics of the vibronic chain consisting of 10 sites, where each site is coupled to $Q=100$ modes ($M=1000$ modes in total). The simulation parameters are the same as Fig.2(c) and (d) in the main manuscript, implying that the reorganization energy is $\lambda=4750$ cm$^{-1}$. The strong environmental coupling makes the electronic excitation localized at the initially populated site 1 due to the strong polaron formation. (b) Population dynamics of a vibronic chain consisting of 10 sites where each site is coupled to $Q=48$ modes, as is the case of Fig.2(d) in the main manuscript. Here the Huang-Rhys factors are reduced compared to the case in the main manuscript in such a way that the overall environmental coupling is the same to the inter-site electronic coupling, $\lambda = J_{n,n+1}$ and the mode damping rate is reduced to $\gamma_q=(1$ ps$)^{-1}$. This is the case of the most non-perturbative regime, and it is found that the convergence can be achieved for $N_b = 5$ and $\chi =12$.}
	\label{fig:SMfurtherres}
\end{figure}

\subsection{Comparison of HEOM and DAMPF}
As is the case of DAMPF, the simulation parameters of HEOM are determined by the multi-exponential fitting of the target BCF \cite{Tanimura1989,Tanimura2006,Ishizaki2009,Lim2018}. Here we compare the simulation cost of these non-perturbative methods for the case that underdamped modes are moderately coupled to electronic states. 

The simulation cost of HEOM is mainly determined by two factors: how many exponentials are needed to fit the BCF and how strongly electronic states are coupled to vibrational modes compared to the mode damping rates. The first factor depends on how many frequency components are needed to fit the BCF, which can be checked by Fourier transforming the BCF from time to frequency domain. At lower temperatures, the BCF can show longer-lived oscillatory features, with the peak structures in the frequency domain becoming narrower. This makes low-temperature HEOM simulations challenging. In addition, even if there are a few underdamped modes, when the vibronic coupling is sufficiently stronger than the mode damping rate, HEOM simulation can become challenging. In HEOM approach, a reduced electronic state is coupled to auxiliary operators, describing system-environment correlations, where the total number of auxiliary operators is unbounded. In practical simulations, one needs to increase the number of auxiliary operators, called tier, until system dynamics shows convergence. Here the vibronic coupling strength determines how quickly the correlations are created, while the mode damping rate governs the decay rate of the correlations. As the vibronic coupling becomes stronger for a given mode damping rate, the number of auxiliary operators required for exact simulations increases, resulting in high memory and simulation time cost.

To compare the simulation cost of HEOM and DAMPF, we consider a dimer system where each site is locally coupled to two Lindblad oscillators, namely $Q=2$ and $M=4$. In this case, the effective BCF of the Lindblad oscillators is described by four exponentials. The first oscillator is modelled by $\omega_1 = 1500$ cm$^{-1}$,  $\gamma_1=($1 ps$)^{-1}$, and $s_1=0.1$ or 0.2. The second oscillator is relatively overdamped, characterized by $\omega_2 = 500$ cm$^{-1}$, $\gamma_2 = ($100 fs$)^{-1}$, $s_2 = 0.1$, dominating electronic dephasing. Both modes are coupled to thermal resorvoirs at room temperature $T_{q=1,2}=300$ K. We assume that both sites have the same energy-level, $\Omega_1=\Omega_2$, with inter-site coupling $J_{12}=500$ cm$^{-1}$. In simulations, a single excitation is localized at site 1 at initial time. Fig.\ref{fig:SMHEOMcomparison} (a) shows the population dynamics of a dimer when the Huang-Rhys factor of the underdamped mode is taken to be $s_1 = 0.1$. It is found that the convergence of HEOM results up to 500 fs is achieved at the 6th tier of hierarchy, which are well matched to the DAMPF results. The number of auxiliary operators within the 6th tier, including reduced electronic state, is $(N N_{\rm exp}+N_{\rm tier})!/(N N_{\rm exp})!/N_{\rm tier}! = 3003$, where $N_{\rm tier}=6$, $N=2$ and$ N_{\rm exp} = 4$, denote, respectively, the tier, the number of sites and the number of exponentials required to fit the BCF of a local bath \cite{Tanimura2006} (see the supplemental material in Lim \textit{et al.} \cite{Lim2018} for more details). The dimension of each auxiliary operator is the same to that of the reduced electronic state, namely a $2\times2$ matrix describing the single excitation subspace. On the other hand, Fig.\ref{fig:SMHEOMcomparison} (b) shows the case when the Huang-Rhys factor is increased to $s_1 = 0.2$. It is found that the convergence of HEOM results up to 450 fs is achieved at the 14-th tier, where the total number of auxiliary operators is 319770, two orders of magnitude larger than the case of $s_1=0.1$. It is notable that HEOM results after 450 fs start to show divergent behavior, implying that the number of auxiliary operators should be increased to obtain exact HEOM results at longer times (confirmed by increasing tier up to 20, not shown here). {We note that HEOM and DAMPF results are also well matched for three site case ($N=3$, not shown here).}

These results demonstrate that the simulation cost of HEOM increases rapidly as the vibronic coupling strength increases. HEOM simulations become more challenging as the number of sites, $N$, and/or the number of exponential terms in the BCF, $N_{\rm exp}$, increases, where $N_{\rm exp}$ depends on how many underdamped modes are present with different vibrational frequencies. Table \ref{table:SMtablememory} shows a comparison of memory cost of HEOM and DAMPF. Importantly, the memory cost of DAMPF increases polynomially, contrary to the factorial growth of HEOM in memory size, enabling one to investigate large vibronic systems coupled to highly structured environments. {The number of elements in a single MPO is $2\chi N_b^2 + (M-2) \chi^2 N_b^2$ and the quantum state is represented by $N^2$ MPOs. In addition, we can reduce the simulation cost by using Hermicity, namely $\hat{\mathcal{O}}_{n,m}=\hat{\mathcal{O}}_{m,n}^\dag$. We note that the matrix product state method is recently applied to HEOM in order to reduce the computational costs \cite{Shi2018}, which is not discussed here.}

\begin{figure}
	\includegraphics[width=0.8\textwidth]{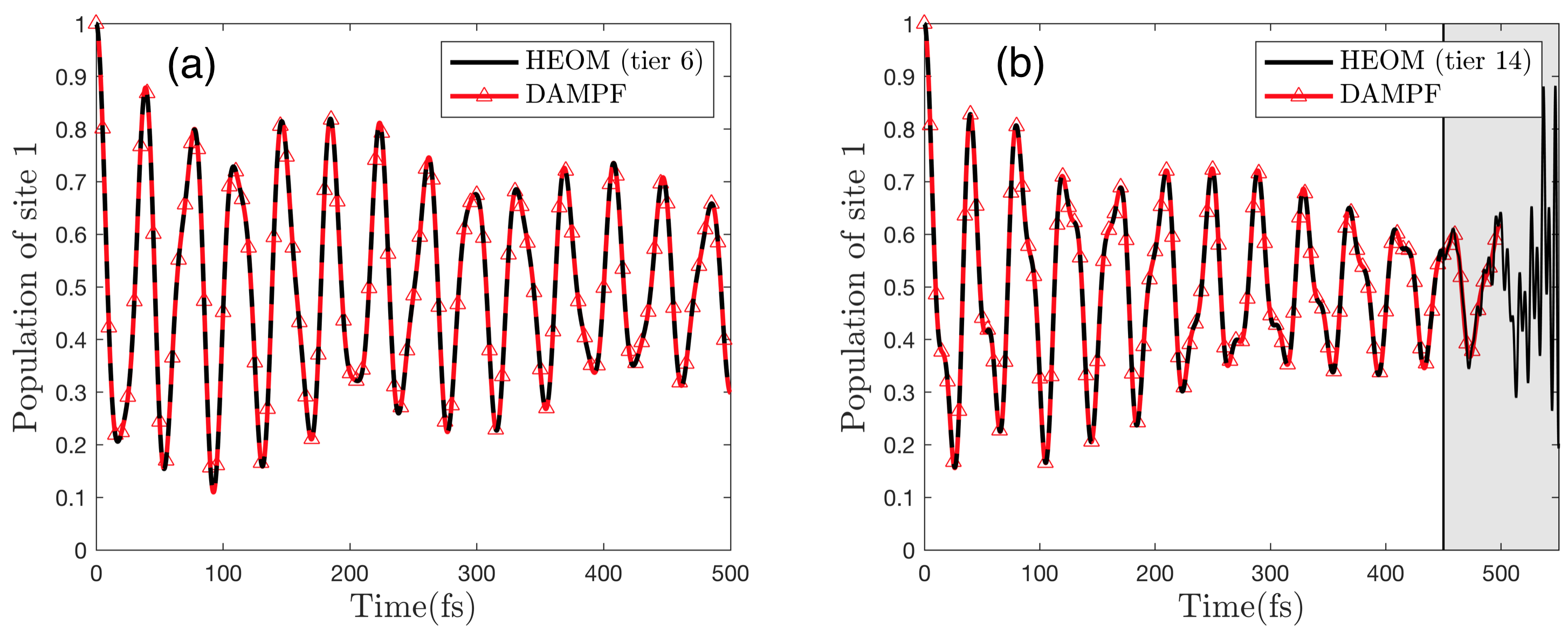}
	\caption{(a) Population dynamics of a vibronic dimer. The Huang-Rhys factor of the underdamped modes is taken to be $s_1=0.$1 (see the main text for the other parameters), where HEOM results show convergence at tier 6. (b) When the vibronic coupling strength is increased to $ s_1=0.2$, HEOM results up to 450 fs show convergence at tier 14. HEOM results after 450 fs show divergent behavior, as shown in a shaded region, where the numerical errors are due to the finite number of auxiliary operators considered in simulations, not due to the low time resolution or other numerical artefacts. Convergence at later times can be achieved by increasing tier further (not shown here). These simulations were carried out with $N_b=8$ levels per oscillator and a bond dimension of $\chi=18$, keeping the local error as defined in \eqref{eq:SMerrorbound} around $10^{-5}$} \label{fig:SMHEOMcomparison}
\end{figure}
\begin{table}
\caption{Dependence of memory cost of HEOM and DAMPF on the number of sites (top, $Q=2$ fixed) and the number of oscillators (bottom, $N=10$ fixed). Note that $Q$ oscillators induce $2Q$ exponential terms in the BCF, including positive and negative frequency components. HEOM simulation cost is based on the assumption that tier is fixed to $N_{\rm tier}=8$, and each auxiliary operator is described by a $N \times N$ matrix (single excitation subspace) with each complex-valued element requiring 16 bytes (double precision). DAMPF simulation cost is estimated based on the assumption that local dimensions are fixed to $N_b=8$ and the bond dimension is $\chi=18$, which are sufficient to obtain exact results for the model parameters considered in this work.}
\begin{ruledtabular}
\begin{tabular}{ccccc}
(Q=2) & N=5    & N=10   & N=15  & N=20   \\ 
\hline
HEOM & 1.25GB & 600 GB & 26 TB & 411 TB \\ 
DAMPF  &   40 MB     &   330 MB     &    1.1 GB   &  2.6 GB \\ 
\end{tabular}
\end{ruledtabular}
\begin{tabular}{ccccc}
 & & & & \\
\end{tabular}
\begin{ruledtabular}
\begin{tabular}{ccccc}
(N=10)             &   Q=2    & Q=12      & Q=24     &Q =48						   \\ 
\hline
HEOM    &   0.6 TB & $10^5$ TB & $10^8$ TB &  $10^{10}$ TB      \\ \hline
DAMPF       &  330 MB   & 2.1 GB    &   4.3 GB  & 8.7 GB \\ 
\end{tabular}
\end{ruledtabular}
\label{table:SMtablememory}
\end{table}

\subsection{Non-classicality of vibrational dynamics}
Semi-classical approaches  circumvent the numerical complexity of quantum harmonic oscillators with the assumption that the oscillators can be well approximated by coherent states. Here we show that for the parameter regime we considered, vibrational dynamics shows non-classicality, implying that the assumption behind semi-classical approaches is not valid. As shown in Ref.\cite{Lemmer2018}, a narrow Lorentzian spectral density can be well described by the effective BCF of an underdamped, high-frequency mode under Lindblad damping. Here we focus on the dynamics of the reduced vibrational state of such a high-frequency mode and investigate its non-classicaility by using the Mandel parameter \cite{Mandel1979}
\begin{align}\label{eq:SMMandel}
\mathcal{M}_m=\frac{\langle\hat{n}_{m}^2\rangle - \langle\hat{n}_m\rangle^2}{\langle\hat{n}_m \rangle }-1
\end{align}
\noindent where $\langle \hat{n}_m\rangle$ and $\langle \hat{n}_m^2\rangle$ are the first and second moments, respectively, of the occupation number operator $\hat{n}_m=\hat{a}_{m,1}^\dag \hat{a}_{m,1}$ of the underdamped mode coupled to site $m$. The Mandel parameter measures the deviation of the occupation number distribution from Poisson distribution. Negativity of $\mathcal{M}_m$ indicates the presence of non-classical features of the $m$-th oscillator. In Fig. \ref{fig:SMmandel_parameter}, the dynamics of the Mandel parameters of underdamped modes are displayed for a vibronic chain consisting of 10 sites ($\Omega_m=\Omega_n$, $J_{n,n+1}=400$ cm$^{-1}$, $\omega_1=1500$ cm$^{-1}$, $s_1 = 0.1$, $\gamma_1=(1$ ps$)^{-1}$, $\omega_2=500$ cm$^{-1}$, $s_2=0.1$, $\gamma_2=(50$ fs$)^{-1}$). The Mandel parameter for the first and second oscillators, shown in black and red, respectively, clearly show negativity. These results demonstrate the presence of non-classical features in vibrational dynamics, as well as the fact that reduced vibrational states can be monitored within our DAMPF approach.

 \begin{figure}
	\includegraphics[width=0.5\textwidth]{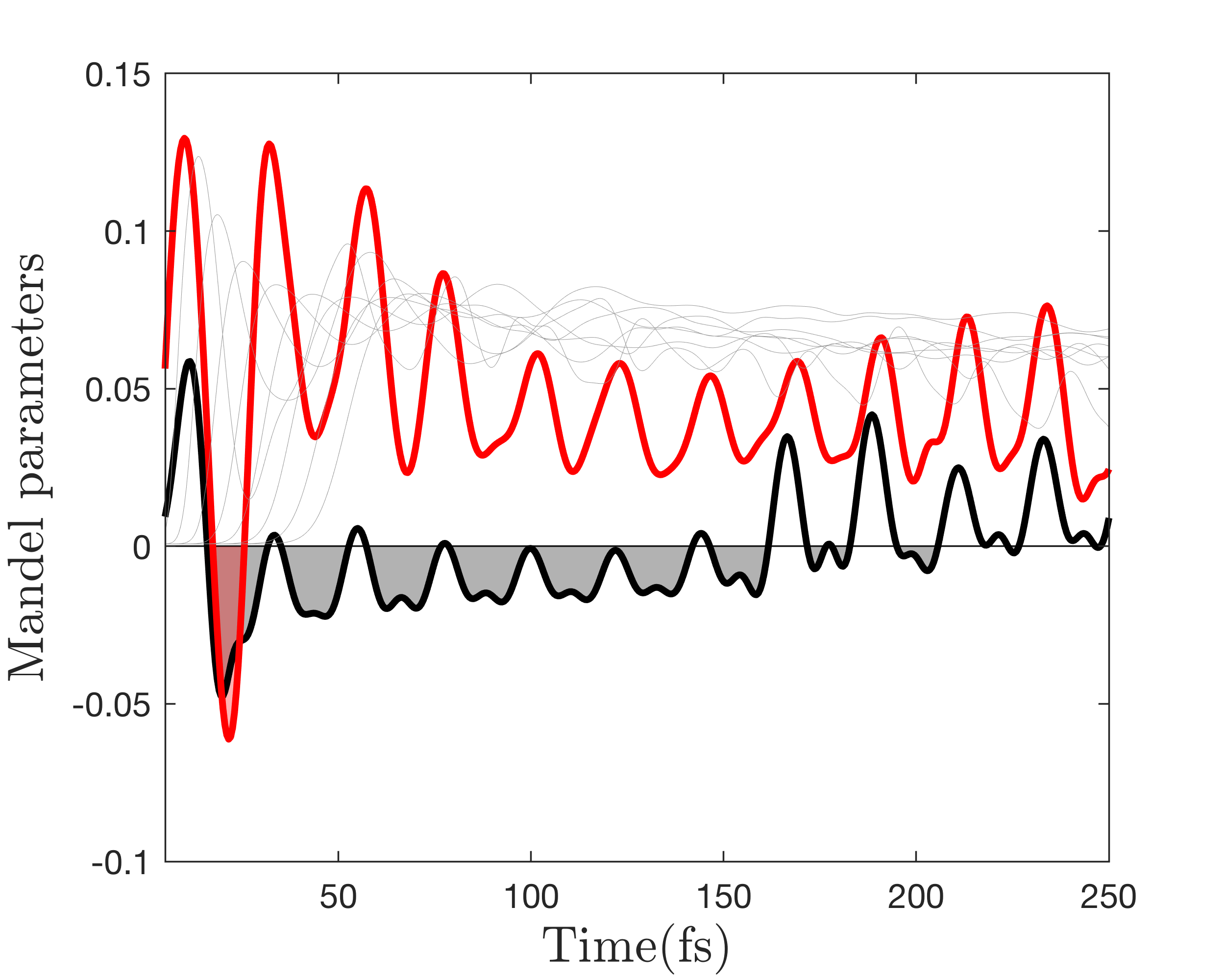} 
\caption{Mandel parameter of underdamped vibrational modes. We consider a vibronic chain consisting of 10 sites  and $J_{n,n+1}=400$ cm$^{-1}$ with high frequency mode $\omega_1= 1500$ cm$^{-1}$, $s_1=0.1$, $\gamma_1=(1$ ps$)^{-1}$, $T_1=300$ K, $N_{b,1}=8$ and the additional overdamped mode $\omega_2=500$ cm$^{-1}$, $S_2=0.1$, $\gamma_2=(50$ fs$)^{-1}$, $T_2=300$ K,$N_{b,2}=4$, where the initial electronic excitation is localized at site 1. The Mandel parameters of the underdamped modes coupled to site 1 and 2 are shown in black and red, respectively. Note that the Mandel parameters show negativity, indicating the non-classicaility of the reduced vibrational states of the two modes. The Mandel parameters of the other eight underdamped modes are shown in grey.}
	\label{fig:SMmandel_parameter}
\end{figure}

\subsection{Signatures of Non-Markovianity}
Non-Markovian effects of environments on system dynamics can be studied by the DAMPF, as it is a numerically exact method. As an example, we consider a vibronic chain consisting of 10 sites, and investigate the time evolution of the reduced electronic state starting from two different initial conditions, namely the lowest and highest energy excitons. Here the exciton states are defined as the eigenstates of the electronic Hamiltonian $\hat{H}_e$ (Eq.\eqref{eq:SMHe}). The distance between two states can be expressed as
\begin{align}
\label{eq:SMtrace_distance}
D(\hat \rho_{e1}(t),\hat \rho_{e2}(t))= \frac{1}{2} \sqrt{(\hat{\rho}_{e1}(t)-\hat{\rho}_{e2}(t))^\dag (\hat{\rho}_{e1}(t)-\hat{\rho}_{e2}(t)) }.
\end{align}
where $\hat{\rho}_{e1}(t)$ and $\hat{\rho}_{e2}(t)$ denote the time-evolved reduced electronic states. In the case of Markovian noise, the distance decays monotonically, as the states converge to the steady state \cite{Rivas2014,Breuer2015}. On the other hand, in the case of non-Markovian noise, the distance can transiently increase in time, which is the signature of non-Markovianity \cite{Breuer2015}. Fig.\ref{fig:SMtracedistance} shows the presence of non-Markovianity for various Lindblad oscillator parameters. A relatively high bond dimension $\chi=25$ was required for numerically exact simulations, due to the correlated dynamics of the oscillators induced by the delocalized initial electronic state.
 \begin{figure} 
	\includegraphics[width=0.5\textwidth]{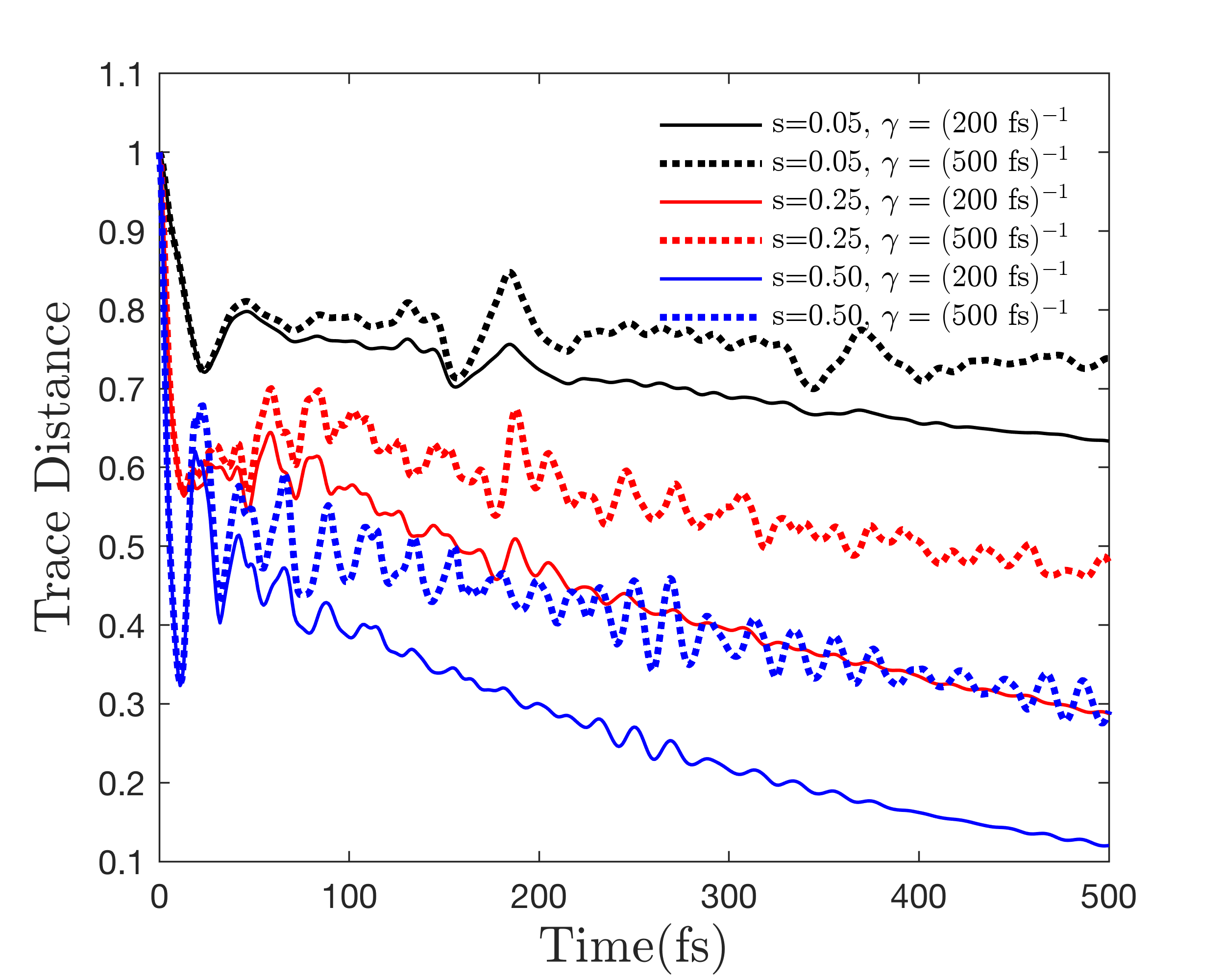} 
	\caption{Trace distance between reduced electronic states starting from lowest and highest energy excitons, respectively. We consider a vibronic chain consisting of 10 sites ($\Omega_m=\Omega_n$, $J_{n,n+1}=500$ cm$^{-1}$), where each site is coupled to a single oscillator modeled by $\omega_1=1500$ cm$^{-1}$, $s_1 \in\{0.05, 0.25\}$, $\gamma_1 \in\{ (200$ fs$)^{-1}$, (500 fs$)^{-1}\}$, $T_1=0$. The temporal increase of the trace distance indicates the presence of non-Markovian effects.} 
	\label{fig:SMtracedistance}
\end{figure}

\end{widetext}

\bibliographystyle{/usr/local/texlive/2016/texmf-dist/bibtex/bst/revtex/apsrev4-1}

\end{document}